\documentclass[sigconf]{acmart}

\usepackage[frozencache,cachedir=.]{minted}

\setcopyright{acmcopyright}
\copyrightyear{2018}
\acmYear{2018}
\acmDOI{10.1145/1122445.1122456}

\acmConference[HILT '20]{Workshop on Safe Languages and Technologies for Structured and Efficient Parallel and Distributed/Cloud Computing}{November 15--20, 2020}{Chicago, IL}
\acmBooktitle{HILT '20: Workshop on Safe Languages and Technologies for Structured and Efficient Parallel and Distributed/Cloud Computing}
\acmPrice{15.00}
\acmISBN{978-1-4503-XXXX-X/18/06}

\begin{document}

\title{Extended Abstract: Productive Parallel Programming with Parsl}

\author{Kyle Chard}
\affiliation{%
  \institution{University of Chicago}
}
\author{Yadu Babuji}
\affiliation{%
  \institution{University of Chicago}
}
\author{Anna Woodard}
\affiliation{%
  \institution{University of Chicago}
}
\author{Ben Clifford}
\affiliation{%
  \institution{University of Chicago}
}
\author{Zhuozhao Li}
\affiliation{%
  \institution{University of Chicago}
}
\author{Mihael Hategan}
\affiliation{%
  \institution{University of Chicago}
}
\author{Ian Foster}
\affiliation{%
  \institution{University of Chicago}
}

\author{Mike Wilde}
\affiliation{%
  \institution{ParallelWorks}
 }

\author{Daniel S. Katz}
\affiliation{%
  \institution{University of Illinois at Urbana-Champaign}
  }

\renewcommand{\shortauthors}{Chard, et al.}

\begin{abstract}
  Parsl is a parallel programming library for Python 
  that aims to make it easy to specify parallelism in 
  programs and to realize that parallelism on arbitrary
  parallel and distributed computing systems. Parsl
  relies on developers annotating Python functions---wrapping either Python or external applications---to indicate that these
  functions may be executed concurrently. Developers can then
  link together functions via the exchange of data. 
  Parsl establishes a dynamic
  dependency graph and sends tasks for execution on connected
  resources when dependencies are resolved. Parsl's runtime system
  enables different compute resources to be used, from laptops
  to supercomputers, without modification to the Parsl program.
\end{abstract}




\maketitle

\section{Introduction}

As we approach the limitations of sequential processing power, 
computer architectures are becoming
increasingly parallel and distributed. 
Unfortunately, parallel and distributed computing has a reputation
for being complex, frail, and unsafe. To address the needs of a diverse
developer community new programming languages, libraries, and tools
are needed to better enable  productive, safe, robust and portable 
parallel and distributed programming. 

Python has established itself as one of the most productive programming languages
as it is easy to use and has a thriving user community and ecosystem of libraries and tools. As a result, Python has been broadly
adopted in industry and academia. However, one of the most well-known
limitations of Python is its use of the Global Intpreter Lock (GIL) 
lock that limits concurrent execution of threads---
and the resulting implications 
with respect to parallelization. 
Overcoming this limitation has been the focus of many
Python libraries, for example, Python's multiprocessing library allows applications to spawn new processes for execution before they are rejoined to the master process upon completion. While multiprocessing
addresses the need for concurrent execution on a node, it does not
support execution in a distributed setting. 

Parsl is a Python library that augments Python to enable productive, safe, robust
and portable parallel and distributed programming. Parsl's productivity
stems from its simple extensions to Python in which developers
express opportunities for concurrent execution using function decorators.
At runtime, Parsl establishes a dynamic dependency
graph comprised of tasks (i.e., calls to Python functions) with edges 
representing shared input/output data between tasks. 
Parsl encodes this information as a Directed Acyclic Graph (DAG), which 
it uses to implement a safe concurrency model in which tasks are only executed
when their dependencies (e.g., input data dependencies) are met. 
When the program executes, Parsl manages the execution of function 
invocations on various computing resources, from laptops to supercomputers.
Parsl tracks task execution, detects exceptions, retries tasks when they fail, and is able 
to overcome various faults (e.g., node failure, task failure). Finally,
to enable programs to be moved between different systems, 
Parsl separates program implementation from runtime configuration thereby
enabling developers to load a system-specific Python configuration object at runtime.

In this extended abstract we highlight Parsl's productive programming model. Further
details of Parsl's implementation and runtime 
model is available in prior publications~\cite{babuji19parsl, babuji19scalable, babuji18parsl}

\section{Parsl Programming Model}

Parsl augments Python with  constructs to enable
specification of parallelism in Python programs. 
Parsl uses these constructs
to establish a dynamic dependency graph via which it 
can determine a safe and portable execution plan. 

\subsection{Parsl Apps}
At the core of the Parsl model are Parsl \textit{apps}---decorated
Python functions that wrap either pure Python code (\texttt{python\_app})
or external applications that can be invoked via the shell (\texttt{bash\_app}).  Listing~\ref{lst:parsl-apps} shows
how Parsl apps can be used to print ``Hello world''.
Parsl apps are executed asynchronously and thus they 
must include all context needed for execution. 
For example, dependencies must be imported in the app
and required data must be explicitly passed via
arguments. The Parsl \texttt{bash\_app} uses the 
return statement to specify the Bash command
to be executed. 

\begin{listing}[h]
  \begin{minted}[frame=lines]{python}
@python_app
def hello():
    return 'Hello world'
    
@bash_app
def hello():
    return 'echo "Hello world"'
    
  \end{minted}
    \vspace{-2ex}

  \caption{Hello world Python and Bash apps.
    \label{lst:parsl-apps}
  }
\end{listing}
 
\subsection{Futures}
As Parsl apps are executed asynchronously, and perhaps on
remote resources with variable delays, it would be inefficient
for the Python program to wait for the app to complete execution.
Instead, Parsl supports concurrent execution as follows. Whenever a Parsl program calls an app, Parsl will create a new task in its dependency
graph and immediately return a future in lieu of that function’s result(s). The program will not block and can continue immediately through the program. At some point, for example when the task’s dependencies are met and there is available computing capacity, Parsl will execute the task. Upon completion, Parsl will set the value of the future to contain the task’s output. Listing~\ref{lst:future} shows an example of the future being 
returned from the invocation of the \texttt{hello} app. Parsl's
futures also provide methods for inspecting the current status
and accessing the result.

\begin{listing}
  \begin{minted}[frame=lines]{python}
@python_app
def hello ():
    import time
    time.sleep(5)
    return 'Hello World!'

app_future = hello()

# Check if the app future is resolved
print('Done: {}'.format(app_future.done()))

# Wait for the future to resolve
print('Result: {}'.format(app_future.result()))
  \end{minted}
    \vspace{-2ex}

  \caption{Invocation of a Parsl app will return a future
  to the calling program. The future can be used to retrieve
  the result when the app completes executing.
    \label{lst:future}
  }
\end{listing}

Parsl allows futures to be passed as input
to other Parsl apps, thereby creating a dependency
between the app that produces the future and the 
app that consumes that future. Parsl monitors
these dependencies and as futures are resolved it determines
what dependent apps may now be
executed.

\subsection{Data}

Parsl supports the exchange of both Python objects and 
external files between Parsl apps. To enable portability, 
and simplify use, Parsl aims to abstract execution location 
by ensuring that apps may access the same input arguments and files irrespective of where the app is executed. 

Listing~\ref{lst:data} illustrates how apps
can communicate using standard Python parameter
passing and return statements.  Parsl enables passing
of primitive types, files, and other complex types 
that can be serialized (e.g., numpy array, scikit-learn model).

\begin{listing}
  \begin{minted}[frame=lines]{python}
@python_app
def communicate(name):
    return 'hello {0}'.format(name)

r = communicate('bob')
print(r.result())
  \end{minted}
    \vspace{-2ex}

  \caption{App communication via Python arguments.
    \label{lst:data}
  }
\end{listing}

Listing~\ref{lst:files} shows how Parsl apps can 
communicate via files. Parsl defines a \texttt{file} object
to abstract file location and relative paths for file access. 
A file may be passed as an input argument to an app or
returned from an app after execution. Parsl's 
data management features support automatic transfer (i.e., staging) of 
files between the main Parsl program, worker nodes, or 
external data storage systems.
Input files can be passed as regular input arguments. 
When executing within an app, the \texttt{filepath} attribute of a File can be used to determine the location of the file on the execution system's file system.
Output file objects must also be specified at app invocation such that Parsl
can track the creation of the file and subsequent staging back to the 
main program or other executing apps. Output files are specified 
with the app's \texttt{outputs} parameter. 

\begin{listing}
  \begin{minted}[frame=lines]{python}
from parsl.data_provider.files import File
    
@python_app
def sort_numbers(in_file):
    with open(in_file.filepath, 'r') as f:
        strs = [n.strip() for n in f.readlines()]
        strs.sort()
        return strs
    
unsorted_file = File(
    'https://raw.githubusercontent.com/Parsl/' + 
    'parsl-tutorial/master/input/unsorted.txt')
    
f = sort_numbers(unsorted_file)
print (f.result())
  \end{minted}
    \vspace{-2ex}
  \caption{App communication via files. In this case a remote file
  is passed to an app that sorts the contents of that file.
    \label{lst:files}
  }
\end{listing}

\subsection{Configuration}
Parsl separates program logic from execution configuration, enabling programs to be developed in a way that is agnostic of execution environment. Configuration is expressed in a Python object (Listing~\ref{lst:config}) which is loaded at runtime. 
The configuration object enables developers to introspect permissible options, validate configurations, and dynamically
modify configurations during execution. 
The configuration specifies details of the provider, executors, connection channel, allocation size, and data management options.

\section{Related Work}

Considerable prior work has explored
methods for supporting parallelism in applications. 
We briefly review methods that are offered as
domain specific languages, as libraries in an existing language, 
and as language-independent frameworks.

There are a number of domain specific languages and 
workflow systems that support the orchestrated
execution of task dependency graphs. 
Systems, such as Pegasus~\cite{pegasus} 
implement a static DAG model in which developers define
the structure of the program in a custom representation
and then they execute it via the workflow system.
Python-based workflow systems such as
FireWorks~\cite{jain2015fireworks}, Apache Airflow~\cite{airflow}, and Luigi~\cite{luigi} provide similar capabilities within Python. 
Swift~\cite{swift} and NextFlow~\cite{di2017nextflow}
implement their own DSL which is evaluated to generate
a DAG.

Most well-known programming languages offer a range of libraries designed
to support parallel and distributed execution.
In Python, Dask~\cite{dask} supports parallel data analytics 
via custom implementation of common Python libraries (e.g., Pandas)
and a general distributed runtime for execution on clusters. 
FaaS systems, such as funcX~\cite{chard20funcx}, often use
similar methods for distributed execution.

Other systems take a language-independent approach to developing
parallel and distributed applications. 
Concurrent Collections~\cite{burke11cnc} implements a language-independent
way of encoding parallelism in different host languages. Developers
identify data and control dependencies, and encode these dependencies in a graph. 
The graph is executed by translating the specification to code for a 
specific runtime system (e.g., in C++, Java, and .NET).
OpenMP~\cite{dagum98openmp} provides a set of language- and platform-independent directives
for augmenting an application and parallelizing execution on nodes.
It is often combined with MPI for distributed execution. 

\begin{listing}
  \begin{minted}[frame=lines]{python}
from parsl.config import Config
from parsl.channels import LocalChannel
from parsl.providers import SlurmProvider
from parsl.executors import HighThroughputExecutor
from parsl.launchers import SrunLauncher
from parsl.addresses import address_by_hostname

config = Config(
    executors=[
        HighThroughputExecutor(
            label="frontera_htex",
            address=address_by_hostname(),
            max_workers=56,
            provider=SlurmProvider(
                channel=LocalChannel(),
                nodes_per_block=128,
                init_blocks=1,
                partition='normal',
                launcher=SrunLauncher(),
            )
        )
    ]
)   
  \end{minted}
  \vspace{-2ex}
  \caption{Parsl configuration for running on TACC's Frontera supercomputer.
    \label{lst:config}
  }
\end{listing}

\section{Summary}
Parsl offers a productive way of implementing portable
parallel and distributed programs in Python. The benefit of extending
Python with simple extensions has enabled a diverse 
range of developers to leverage
Parsl in various domains and use cases.
The modular configuration and execution model allows 
Parsl programs to be moved between different parallel
and distributed computing environments.
In prior work we have shown that Parsl can execute millions of tasks, 
scale to more than 250,000 workers across more than
8000 nodes, and process upward of 1200 tasks per second~\cite{babuji19parsl}.

\begin{acks}
Parsl is supported by NSF award ACI-1550588
and DOE contract DE-AC02-06CH11357. 
\end{acks}

\bibliographystyle{ACM-Reference-Format}
\bibliography{refs}


\end{document}